\documentclass{PoS}
\newcommand{\dcp}{\delta_{CP}}
\newcommand{\nova}{NO$\nu$A\ }

\title{What antineutrinos can tell about octant and $\delta_{CP}$ in DUNE? }

\ShortTitle{Antineutrinos role in DUNE}

\author{\speaker{Newton Nath} \\
       Physical Research Laboratory, Navrangpura, Ahmedabad--380 009, India, \\
       Indian Institute of Technology, Gandhinagar, Ahmedabad--382 424, India\\
        E-mail: \email{newton@prl.res.in}}

\author{Monojit Ghosh\\
        Physical Research Laboratory, Navrangpura, Ahmedabad--380 009, India \\}

\author{Srubabati Goswami\\
        Physical Research Laboratory, Navrangpura, Ahmedabad--380 009, India}

\abstract{We study  the efficiency of DUNE, a next generation long baseline oscillation experiment to resolve two major unknowns in neutrino oscillation physics. These are, octant of $\theta_{23}$ (i.e. if $\theta_{23}$ is $< 45^\circ$ or $>45^\circ$) and Dirac CP phase $\delta_{CP}$. We mainly focus on the role of antineutrinos when they travel 1300 km baseline of DUNE.
We observe that for DUNE, the antineutrino runs help to remove parameter degeneracies even in the parameter space where the antineutrino probability suffers from various degeneracies. We study these points in detail and find that, due to enhanced  matter effect longer baseline experiments create an increased tension between the 
neutrino and the antineutrino probabilities which helps to increase total sensitivity in case of combined runs.
 We also find that,  antineutrino run increases overall CP sensitivity due to its ability to abolish octant-$\delta_{CP}$ degeneracy.}

\FullConference{38th International Conference on High Energy Physics\\
		3-10 August 2016\\
		Chicago, USA}

\begin{document}

\section{Introduction}
The standard three flavor neutrino oscillation scenario consists of  two mass squared differences ($\Delta m^2_{i1}, i=2,3$) 
%which control the oscillations of the solar and the atmospheric neutrinos respectively
, three mixing angles ($\theta_{ij}, j>i=1,2,3$)  and the Dirac CP phase $\delta_{CP}$. Various neutrino oscillation experiments have measured or given a hints about these parameters. The CP phase $\delta_{CP}$ remains one of the least known parameters among all these.  Other than this, what we do not know in neutrino oscillation physics are (i) the sign of $|\Delta m^2_{31}|$($\Delta m^2_{31} > 0$ is known as the normal hierarchy (NH) or $\Delta m^2_{31} < 0$ is  known as the inverted hierarchy (IH)),  (ii) the octant of $\theta_{23}$( $\theta_{23} < 45^\circ$ is known as the lower octant (LO) or $\theta_{23} > 45^\circ$ is  known as the higher octant (HO)). The current best fit values and their $3\sigma$ ranges are given in  \cite{Gonzalez-Garcia:2014bfa,global_valle} by performing global analysis of neutrino oscillation data. Ongoing long baseline oscillation experiments like, T2K
 \cite{t2k_dcphierocthint} and \nova \cite{Adamson:2016tbq} 
can give information on these unknown parameters. The main difficulties which these experiments have to overcome are the problem of parameter degeneracies i.e. different parameter sets giving equally good fit to the experimental data. 

In this work, we focus on the determination of octant degeneracy and the Dirac CP phase $\delta_{CP}$. The primary channel to determine octant of $ \theta_{23} $  is the disappearance channel $ P_{\mu \mu} $, but for shorter baseline experiments, it suffers from intrinsic octant degeneracy i.e. for $\theta_{23} $ and $ 90^\circ - \theta_{23}  $ one gets the same probability value. Whereas appearance channel $ P_{\mu e} $ does not suffer from intrinsic degeneracy because of the  $\sin^2 \theta_{23} \sin^2 2\theta_{13}$ dependency, but suffer from generalized degeneracies as pointed out in \cite{Barger:2001yr, Minakata:2001qm, BurguetCastell:2002qx, Coloma:2014kca, Ghosh:2015ena}. Therefore, we focus on the combined run of both the channels which can be helpful to determine octant of $ \theta_{23} $ because of their different functional dependency on $ \theta_{23} $. We also explore the non-trivial contribution of antineutrino run in enhancing CP sensitivity by considering hierarchy and octant as known, as well as unknown.
% Also, we examince the role of antineutrino for enhancing octant sensitivity and CP violation (CPV) discovery potential of 1300 km baseline experiment like, DUNE.
In our analysis, we consider 10 kt liquid argon detector and 1.2 MW beam for DUNE \cite{DUNE} and the remaining details that we have considered in this work are given in \cite{Nath:2015kjg} and the references there in. 
\section{Results}
In this section, we present the octant sensitivity and CP violation (CPV) discovery potential of DUNE. We have four possible combinations of (hierarchy$ - $octant) depending on true hierarchy and true octant, namely: NH-LO, NH-HO, IH-LO and IH-HO. Here, we only discuss the cases NH-LO and NH-HO. The detailed descriptions of all the four cases are described in \cite{Nath:2015kjg}.

\begin{figure}
%\vspace{-1.7cm} 
%        \begin{tabular}{lr}
%               \hspace*{0.15in} 
                \includegraphics[width=.6\textwidth]{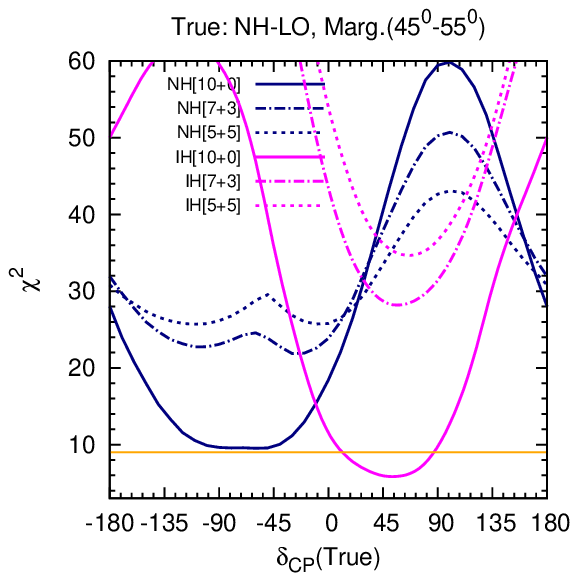}
%                & 
               \hspace*{-0.9in}
               \includegraphics[width=.6\textwidth]{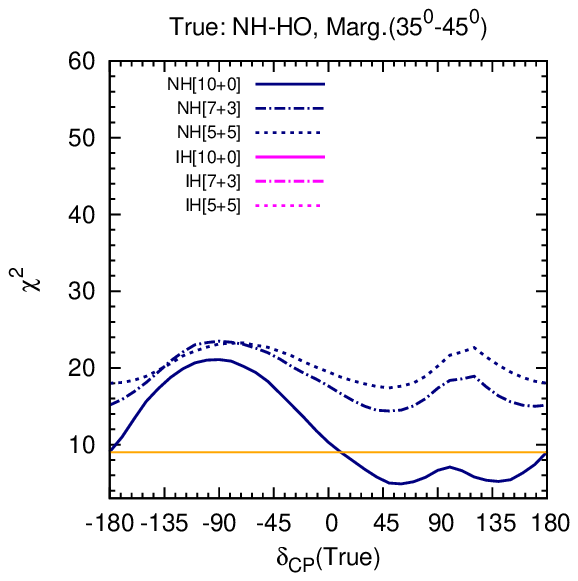}
%        \end{tabular}
\vspace{-0.6cm}        
\caption{Octant discovery $ \chi^{2} $ for DUNE, considering true hierarchy as NH. Here, left (right) column is for true $ \theta_{23}$ = $39^\circ(51^\circ)$  and test($ \theta_{23}$) is marginalized over opposite octant.}
\label{fig:oct_discovery}
\end{figure}

In figure [\ref{fig:oct_discovery}], we present the octant sensitivity of DUNE.  In both the columns, dark-blue (magenta) curves represents True:NH - Test:NH (True:NH - Test:IH). The horizontal yellow line represents the octant sensitivity at 3$ \sigma $ C.L.  True values of $ \theta_{23} $ for the two columns are considered as, 39$^\circ$ (left) and $ 51^\circ$ (right) respectively.
%%%%%%%%%%%%%%%%%%%%%%%%%%%%%%%%%%%%%%%%%%%%
%\vspace{-0.4cm}
%$$$$$$$$$$$$$$$$$$$$$$$$$$$$$$$
From the left column, we can see that DUNE can reach 3$ \sigma $ octant sensitivity with pure neutrino run i.e. [10+0] year for known hierarchy (see solid blue curve).
% Also, in the upper half plane (UHP) i.e. $0^\circ < \dcp < 180^\circ$, only neutrino data provides a better octant sensitivity.
However, for unknown hierarchy we see that only neutrino run is not able to resolve the octant degeneracy even at 3$ \sigma $ due to the wrong hierarchy(WH)-wrong octant(WO) solutions appearing in the region $ 9^\circ < \dcp < 90^\circ $(see solid magenta curve).
% The main point to notice here is that,  though only neutrino run suffers from octant degeneracy, still we get octant sensitivity around 3$ \sigma $. This is one of the special features of the on axis beam based neutrino oscillation experiment where the degeneracy does not exist over the complete energy range and  one can still have  some octant sensitivity only from the neutrino channel. However, storyline changes  if the hierarchy is unknown then there is a   wrong hierarchy(WH)-wrong octant(WO) solutions in the region $ 9^\circ < \dcp < 90^\circ $.  
We find that once antineutrino run is added with neutrino, considering the case with (7+3) years of ($ \nu + \overline{\nu}  $) run, octant degeneracy can be resolved with more than 4$ \sigma $ C.L.  without any information of the true hierarchy for any value of $\dcp$.
The right column describes the case for true (NH-HO). We see that there is 3$ \sigma $ octant sensitivity in the lower half plane (LHP) i.e. $-180^\circ < \dcp < 0^\circ$ even with only neutrino run whereas upper  half plane (UHP) i.e. $0^\circ < \dcp < 180^\circ$ suffers from octant degeneracy in neutrino mode. But with the (5+5) years of ($ \nu + \overline{\nu}  $) run more than 4$\sigma$ octant sensitivity can be attained\footnote{Note that in the rhs of fig.(\ref{fig:oct_discovery}) does not have any magenta curve because of the absence of WH-WO solution for NH-HO.}.
% It is seen that for IH, because of the enhancement of the antineutrino probability due to  matter effect,
%  a large octant sensitive contribution to the $\chi^2$ is added. These combinations of hierarchy$ - $octant can resolve octant degeneracy with  5$ \sigma $ sensitivity with (5+5) years of ($ \nu + \overline{\nu} $) run for any value of $ \delta_{CP} $.
%%%%%%%%%%%%%%%%%%%%%%%%%%%%%%%%%%%%%%%%%%%%%%%%%%%%%%%%%%%%%%%%%%%%%%%%%
%/******************************************************
\begin{figure}
%\vspace{-1.5cm}
%        \begin{tabular}{lr}
%               \hspace*{0.65in}
       \includegraphics[width=0.6\textwidth]{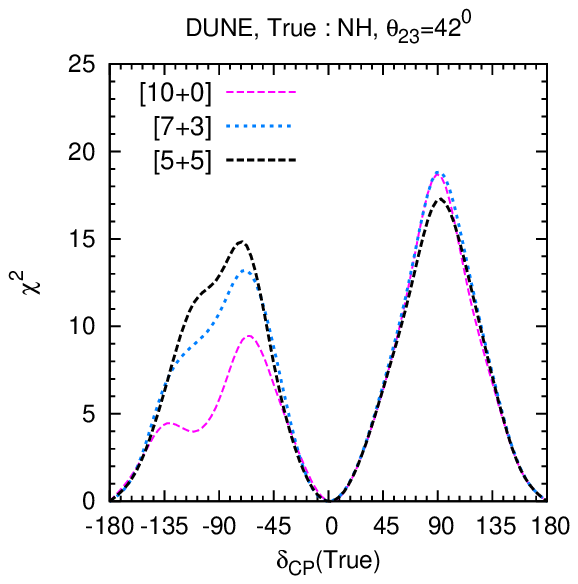}
       \hspace*{-0.9in}
        \includegraphics[width=0.6\textwidth]{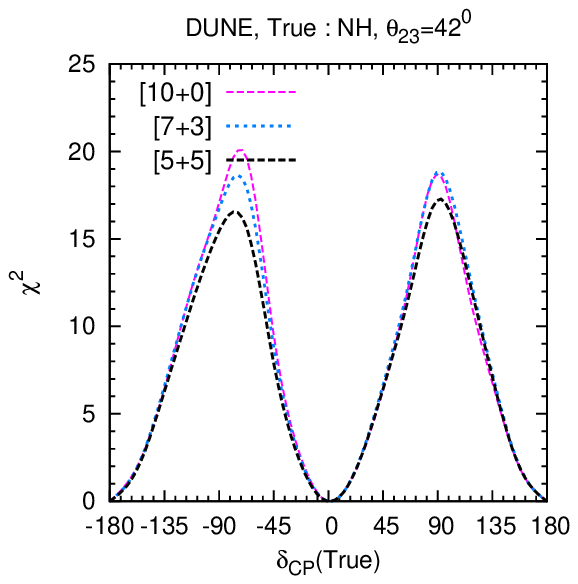}
%        \end{tabular}
\vspace{-0.7cm}
\caption{Both the columns represent CPV $ \chi^{2} $ for unknown(known) hierarchy and octant respectively.}
\label{dune_cp_1}
\end{figure}
\vspace{-0.3cm}
%$$$$$$$$$$$$$$$$$$$$$$$$$$$$$$$$$$$$$$$$$$$$$$$$$$$$$$$$$$$$$$ 

In left (right) column of figure [\ref{dune_cp_1}], we present the CP discovery $\chi^2$ as 
a function of true $\dcp$ for unknown(known) hierarchy and octant. The CPV discovery potential of an experiment is defined  as to distinguish between CP ($\dcp \neq 0^\circ,180^\circ$) violating values from CP conserving values. This figures show the role of antineutrinos
in discovering CP phase and the dependency  with  the octant of $ \theta_{23} $. 
%The antineutrino run plays an important role for LO  near true $\dcp = -90^\circ$
%for the case with unknown hierarchy and octant. 
Comparing left column with right column we see that the CP sensitivity for -90$^\circ$-LO improves for known hierarchy and octant, infact 10+0 case provides
%We observe from right column that for  -90$^\circ$-LO, 10+0 gives
the best sensitivity. We observe that this enhancement for known hierarchy and octant is also true for IH. This indeed proves the fact that, antineutrino run is an instrumental for the removal of the wrong octant solutions.

\begin{figure}
%\vspace{-1.5cm}
%        \begin{tabular}{lr}
%               \hspace*{-0.4in}
                \includegraphics[width=0.6\textwidth]{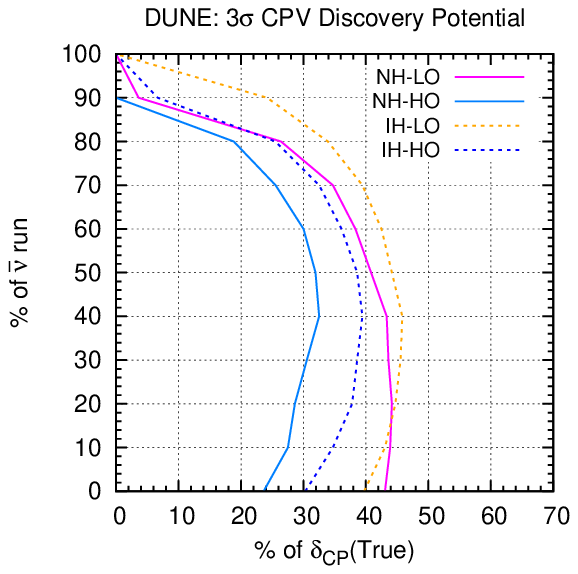}
                \hspace*{-0.9in}
                \includegraphics[width=0.6\textwidth]{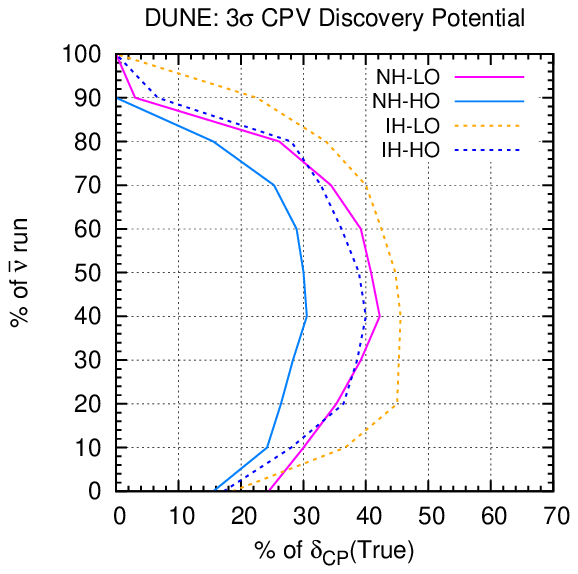}
%        \end{tabular}
\vspace{-0.7cm}
\caption{ CPV discovery plot in  (\% of $ \delta_{CP}(True) $,  \% of antineutrino run) plane at 3$ \sigma $ C.L. First (second) column are for known (unknown) hierarchy and octant.}
\label{dune_cpv_frac}
\end{figure}

In figure [\ref{dune_cpv_frac}], we plot the CPV  discovery potential at $3 \sigma$ C.L. in (percentage of $\overline{\nu} $ run - percentage of $\dcp$) plane. We consider all the four combinations of hierarchy-octant. 
We observe from both the columns that a lesser CP fraction is achieved with the dominant neutrino or antineutrino run.  Comparing both the plots, we find that  maximum CPV fraction can be attained for IH-HO and minimum for NH-HO. Also, 40\% antineutrino run seems to be optimum to achieve maximum CPV fraction in all the cases. 
%%%%%%%%%%%%%%%%%%%%%%%%%%%%%%%%%%%%%%%%%%%%%%%%%%%%%%%%%%%%%%%%%%%%%%%%%
%/******************************************************
\vspace{-0.3cm}
%%%%%%%%%%%%%%%%%%%%%%%%%%%%%%%%%%%%%%%%%%%%%%%%%%%%%%%%%%%

In conclusion, we have examined  the significant role of antineutrinos in 
providing an enhanced  octant and CP sensitivity in the next generation superbeam experiment like DUNE. In our study, we observe that 
%although for some specific parameters only neutrino run is able to  provide $3\sigma$ octant sensitivity for 10 kt detector mass, overall 
combined ($ \nu + \overline{\nu}  $) run gives better sensitivity. In the case of CPV discovery, antineutrino run  plays a leading role due to the synergistic behaviours between neutrinos and antineutrinos even under the assumption of known octant. 
%\section{Acknowledgement}
%Author likes to thank Srubabati Goswami for her valuable suggestions during the preparation of the article.
 
%%%%%%%%%%%%%%%%%%%%%%%%%%%%%%%%%%%%%%%%%%%%%%%%%%%%%%%%%%%%%%%%%%%%%%%%%

\end{document}